%% file: p2_25.tex
\documentclass[aps,twocolumn,pra]{revtex4}

\usepackage{graphicx} 
\usepackage{dcolumn}  
\usepackage{bm}       
\usepackage{amssymb}  
\usepackage{amsmath}  
\usepackage[usenames]{color} 	  

\input epsf

\input dft

\input rotate

\def\bei{\begin{itemize}}
\def\eei{\end{itemize}}
\def\F{_{\sss F}}

\def\xx{_{\sss XX}}

\begin{document}

\title{Non-empirical `derivation' of B88 exchange functional}

\author{Peter Elliott}
\affiliation{Department of Physics and Astronomy, University of California, Irvine, CA 92697, USA}
\author{Kieron Burke}
\affiliation{Department of Chemistry, University of California, Irvine, CA 92697, USA \\ Department of Physics and Astronomy, University of California, Irvine, CA 92697, USA}


\begin{abstract}
The B$88$ exchange energy density functional (created by Becke in 1988) is a crucial part
of the most popular density functional in use today, B$3$LYP.
B88 contains one empirical parameter which was fitted to Hartree-Fock exchange 
energies for the noble gas atoms. We show how local approximations to exchange become
relatively exact under a very specific approach to
the limit of large numbers, but that the usual gradient expansion does not.
The leading corrections can be captured by generalized gradient approximations,
producing a non-empirical derivation of the parameter in B88.
\end{abstract}


\date{\today}

\maketitle

\section{Introduction}
Density functional theory (DFT) has become the method of choice for many electronic structure calculations in quantum chemistry\cite{FNM03}. It balances the demands of accuracy and computation time, making it advantageous to other available methods. Although DFT is a formally rigorous theory, in practice it requires an approximation to the exchange-correlation (XC) energy, $E\xc$, as a functional of the density. In the late $1980$'s and early $1990$'s, approximations, such as B$88$, were developed which proved accurate enough for their application in many areas of both chemistry and physics. The explosion in the use of DFT, driven by newly available computational power, could not have succeeded without the development of such functionals.

Modern DFT began with the Hohenberg-Kohn theorem of $1964$\cite{HK64} and the
Kohn-Sham scheme of $1965$\cite{KS65}. The first approximation to the XC energy
was the local density approximation (LDA), which uses only the
density at a given point to determine the energy density at that point.
This is exact for a uniform gas, since it is the known energy density 
of the uniform gas that is used to define LDA.
The natural successor to LDA is
a semi-local (or gradient-corrected) approximation which adds information about the derivative of the density at that point.
In fact, in the same paper in which LDA is introduced, so too is
the gradient expansion approximation (GEA) for XC.
The coefficients of the GEA are determined by the energy of a slowly-varying gas\cite{AK85,MB68,RG86}.
However it was found that the GEA often worsened LDA results and two decades
 passed before substantial improvements were made.

Generalized gradient approximations (GGAs) effectively resum the gradient
expansion, but using only $|\nabla n|$. The B$88$ functional\cite{B88}
is the most used GGA for exchange overall (as part of B$3$LYP\cite{LYP88,B93}), but the most popular GGA
in solid state applications is PBE\cite{PBE96}.
Neither reduces to the GEA in the limit of small gradients. In this paper we explain
the reason why this must be the case. 
Asymptotic expressions for the energy components as functionals of $N$,
the number of electrons, display `unreasonable accuracy'\cite{S80} even for low $N$.
In order to give good energies for finite systems, any approximate XC functional
must have accurate coefficients in its large-$N$ expansion.
LDA gives the dominant contribution, but the GEA does not yield an accurate
leading correction for atoms.  Popular GGAs such as B88 and PBE do get this correction right.

In Ref. \onlinecite{PCSB06}, the underlying ideas behind this work were developed, however the reasoning was based upon scaling the density and not on the potential scaling discussed below. We refine these ideas and explicitly show how they can be used for functional development, and in particular we show how the parameter in B$88$ may be derived in a non-empirical manner.

\section{Background Theory}
In this section we review the theory behind asymptotic expansions of the energy, including the expression for the  exchange energy. A demonstration of the usefulness of this asymptotic expansion for exchange is also given, where the constant in LDA is found without referring to the electron gas. We also show the form of the gradient correction to LDA exchange, as well as discussing the form of the GGAs, B$88$ and PBE.

\subsection{Asymptotic expansion in $N$}
\label{s:pscal}
Begin with any system (atom, molecule, cluster, or solid)
containing $N$ electrons.   We then imagine changing the number of electrons to 
$N'$.  Since we usually begin from a neutral system, usually we consider only
$N' > N$.  Thus we define a scaling parameter $\zeta=N'/N > 1$.  As we change the
particle number, we simultaneously change the one-body potential $v\ext(\br)$
in such a way as to retain overall charge neutrality, which means
\ben
v\ext^\zeta(\br) = \zeta^{4/3}\, v\ext(\zeta^{1/3} \br),~~~~~~N\to\zeta N.
\label{vzeta}
\een
We refer to this as charge-neutral (CN) scaling.  For an isolated atom, 
$Z\to\zeta Z$ under this scaling, so it remains neutral as the electron
number grows.
For molecules with nuclear positions ${\bf R}_\alpha$
and charges $Z_\alpha$, $Z_\alpha\to \zeta Z_\alpha$
and ${\bf R}_\alpha \to \zeta^{-1/3} {\bf R}_\alpha$. 
In the special case of neutral atoms, the resulting series for the energy is
well-known:
\ben
\label{ESchwing}
E = -a_0 N^{7/3} - a_1 N^{2} - a_2 N^{5/3} - ...
\een
where $a_0 = 0.768745$, $a_1 = -1/2$, and $a_2 = 0.269900$ 
\cite{S80,ES85}.
We say an approximation is large-$N$ \emph{asymptotically exact} to the $p$-th degree (AEp)
if it recovers exactly the first $p+1$ coefficients for a given quantity
under the potential scaling of Eq. (\ref{vzeta}). 
Lieb and Simon\cite{LS73,L81} showed that Thomas-Fermi (TF) theory
becomes {\em exact} in the limit $\zeta\to\infty$ for {\em all} systems.
TF is exact 
in a statistical sense, in that TF gives the correct first term of Eq. (\ref{ESchwing}),
but not the other terms. We say TF is AE0 for the total energy.

A similar expression exists for the exchange component of the energy alone:
\ben
\label{ExSchwing}
E\x = -c_0 N^{5/3} - c_1 N - ...
\een
where $c_0 = 0.2208 = 9a_2/11$ and $c_1$ will be the main topic of this paper.
In a similar fashion, Schwinger demonstrated that the local approximation for
exchange is AE0, and this coefficient is given exactly by local exchange
evaluated on the TF density\cite{S80,ES85,ELCB08}.
However in order to give atomic exchange energies needed for chemical
accuracy, any exchange approximation should be at least AE1.

Now suppose we want to make a local approximation for $E\x$ but know nothing about the
uniform gas.  Dimensional analysis (coordinate scaling\cite{LP85})
tells us that it must be of the form:
\ben
\label{ExLDA}
E\x\LDA[n] = A\x I,~~~~~  I = \int d^3 r \ n^{4/3}(\br).
\een
Requiring that this gives the leading contribution to Eq. (\ref{ExSchwing})
then fixes the value of the constant $A\x$. 
Using any (all-electron) Hartree-Fock atomic code, such as were already available in
the 1960's\cite{F77}, 
one calculates $I$ for densities running down a
particular column of the periodic table
and then deduces its dependence on $Z^{5/3}$.
A modern alternative is to use 
the fully
numerical OPMKS code\cite{Engel} using the OEP exact exchange functional to
find densities for neutral atoms from $Z=1$ to $Z=88$.
By fitting, one finds $I=0.2965 Z^{5/3}$ and hence $A\x = -0.7446$. 
This is remarkably close to the derived result of 
\ben
\label{AxLDA}
A\x = -\frac{3}{4}\left[\frac{3}{\pi}\right]^{1/3} = -0.7386.
\een
Thus, without any recourse to the uniform electron gas, 
we have derived the correct local approximation to $E\x[n]$.\\
This demonstrates that, via asymptotic exactness, the local approximation to exchange
is a universal feature of all systems as $N\to\infty$, when scaled appropriately.
(In fact, Schwinger only proved this for atoms\cite{S80}, we know of no proof for arbitrary systems).

\subsection{Gradient expansions}

The next step up the ladder of increasingly sophisticated density-functional
approximations\cite{PS01} is a
semi-local approximation for $E\x[n]$,
i.e., one which includes information about the gradient of the density.
We introduce the dimensionless measure of the gradient:
\ben
\label{s}
s(\br) = \frac{|\nabla n(\br)|}{2k\F(\br) n(\br)} 
\een
where $k\F(\br) = (3\pi^2 n(\br))^{1/3}$ is the local Fermi wavevector.
This is often written in terms of $x=
{|\nabla n|^2}/{n^{4/3}}$, which is simply proportional to $s$.
Assuming smoothness in $s$ and no preferred spatial direction, we know any sensible
approximation depends only on $s^2$.
The gradient expansion is defined as the expansion of the energy as a functional
of the density around the uniform limit.  The leading correction for exchange is:
\ben
\label{Ex2}
E\x^{(2)}[\n] = \mu \int d^3r\, s^2(\br)\, \epsilon\x\LDA(\n(\br)),
\een
where $\epsilon\x = A\x\n^{4/3}$ and $\mu$ is a constant.
Alternatively, we may write:
\ben
\label{Ex2b}
E\x^{(2)}[\n] =
- \beta\int d^3 r \ n^{4/3}(r) \ x^2.
\een
with
\ben
\label{mubeta}
\beta =  \frac{3}{16\pi}\left[ \frac{1}{3\pi^2}\right]^{1/3}\mu.
\een
In a very slowly-varying electron gas, the gradient is very small,
and the exchange energy will be accurately given by $E\x\LDA+E\x^{(2)}$.
For such systems, the
constant $\mu=10/81$\cite{AK85}, so that $\beta \approx 0.0024$.

The gradient expansion approximation (GEA) means applying this form to
a finite system, using the value of $\mu$ from the slowly-varying gas.  
The GEA for exchange typically reduces the LDA error by about $50\%$.
However it's counterpart for correlation worsens the LDA error, as its energy
density is not even always negative.
In many cases, GEA strongly overcorrects LDA leading to
positive correlation energies and giving poor total energies\cite{MB68}.

\begin{figure}[htb]
\unitlength1cm
\begin{picture}(12,6.5)
\put(-6.2,-3){\makebox(12,6.5){
\includegraphics{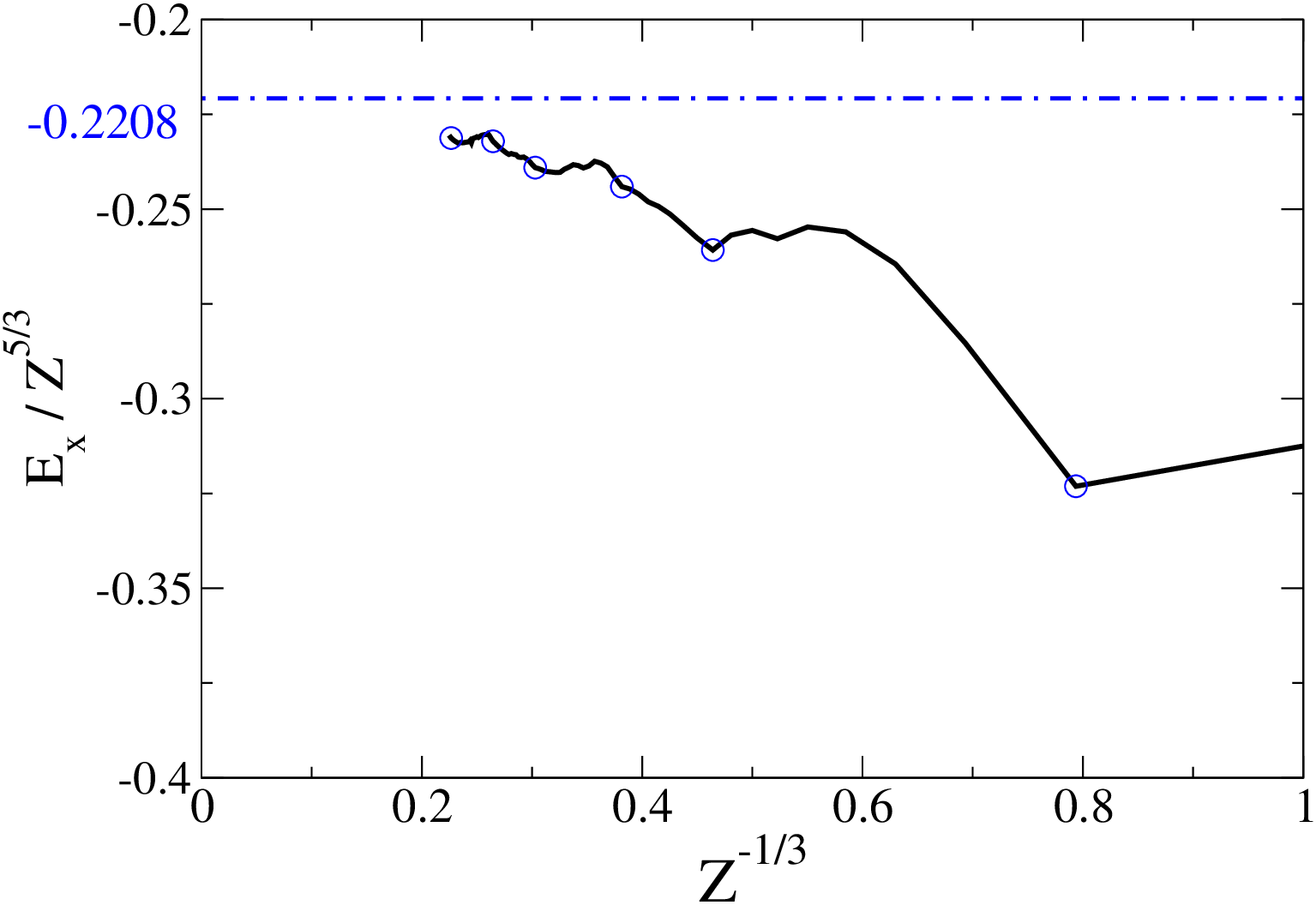}
}}
\end{picture}
\caption{The OEP exact exchange energies E$\xx$ for neutral atoms from $Z = 1$ to $88$, divided by $Z^{5/3}$ in order to pick out the leading term in its asymptotic series. The leading corrections are proportional to powers of $Z^{-1/3}$. The values for the noble gas atoms are given as the circle symbols.}
\label{f:EXX}
\end{figure}

A generalized gradient approximation (GGA) seeks to include
the information contained in $s(\br)$ while improving on the success
of LDA. The B88 exchange functional was designed to reduce to the GEA
form when $s$ is small, but also recover the
correct $-n(r)/2r$ decay of the exchange energy density for large $r$ in atoms.
Thus it interpolates between two known limits, and has the form:
\ben
\label{B88}
\Delta E\x^{\sss B88}[n] 
=- \beta^{\sss B88}\int d^3 r \ n^{4/3}(r) \frac{x^2}{1+6x\beta^{\sss B88}\sinh^{-1}[2^{1/3}x]},
\een 
where $\Delta E\x$ denotes the correction to LDA.
Thus the B$88$ functional\cite{B88} contains one unknown parameter, $\beta^{\sss B88}$. 
In $1988$ Becke found this parameter by
fitting to the Hartree-Fock exchange energies of the noble gases,
finding a value of $0.0053$. In fact, Becke notes that this value is consistent with the observation of a high-$Z$ asymptote for $\beta$. In Ref. \onlinecite{B86}, Becke calculates what value of $\beta$ in Eq. (\ref{Ex2b}) is required in order to give the HF exchange energy for each atom in the first two rows of the periodic table along with the noble gas atoms. Thus, $\beta$ is treated as a function of $Z$, and he observes that it converges for high-$Z$. Thanks to the previous section on asymptotic series, we can now understand why this convergence occurs.
Although the B88 form reduces to that of the gradient expansion for small
gradients, the value for $\beta$ is about twice as large as that predicted from
the slowly-varying gas.

\begin{figure}[htb]
\unitlength1cm
\begin{picture}(12,6.5)
\put(-6.2,-3){\makebox(12,6.5){
\includegraphics{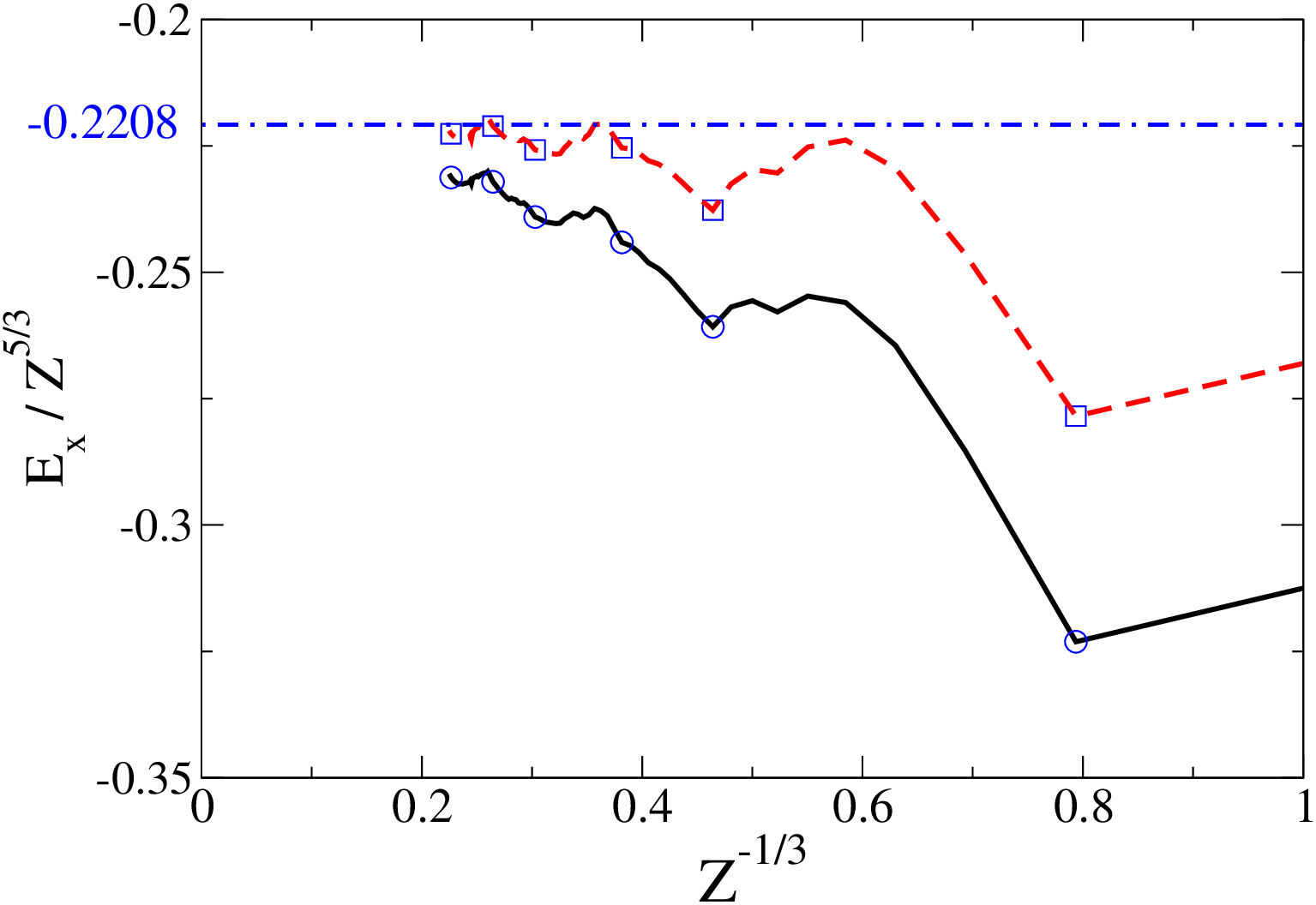}
}}
\end{picture}
\caption{We add to Fig. \ref{f:EXX} the results for the LDA functional evaluated on the OEP densities (dashed-line) keeping the XX values (solid line). As in Fig. \ref{f:EXX}, the Noble gas atom are highlighted with circle and square symbols for XX and LDA respectively.}
\label{f:LDA}
\end{figure}

Another common GGA for exchange is the Perdew-Burke-Ernzerhof (PBE) approximation\cite{PBE96}, usually written
in terms of an enhancement factor, $F\x(s)$, to the LDA exchange energy density:
\ben
E\x^{PBE}[n] = \int d^3 r \ F\x\PBE(s) \ \epsilon\x\LDA[n] 
\een
where
\ben
F\x^{PBE}(s) = 1 + \kappa - \frac{\kappa}{1+\mu s^2/\kappa} ,
\een
and $\mu = 0.2195$ and $\kappa = 0.8040$. 
This form for the enhancement factor is chosen so that it reduces to
LDA for $s=0$ and again recovers the form of the gradient expansion for small $s$.
For large $s$ it becomes a constant determined by the parameter $\kappa$. 
Both $\kappa$ and $\mu$ are determined via satisfaction of various exact conditions. 
The value of $\mu$ was chosen to preserve the good linear
response of LDA for the uniform electron gas under a weak
perturbation\cite{BSA94,MCS95}, while $\kappa$ is set by the 
Lieb-Oxford bound\cite{LO81} on the exchange-correlation energy.  (That condition
is obviously violated by B88, while PBE does not accurately recover the
X energy density in the tails of Coulombic systems).

\section{Theory}

LDA yields the dominant term in 
{\em either} the asymptotic charge-neutral expansion ($\zeta\to\infty$)
{\em or} the gradient expansion for the slowly-varying electron gas, $s\to 0$.
We next show that, contrary to popular myth, the important expansion is the
charge-neutral expansion, {\em not} the gradient expansion.  

The charge-neutral expansion can be applied to any type of matter, be it molecule
or extended solid.  For any finite system, the density decays exponentially 
far from the nuclei.  This is a key distinction between finite systems and
bulk matter, treated with periodic boundary conditions.  Bulk matter has
no such regions.

But, for slowly-varying gases, or more generally when
there are no classical turning points at the Fermi surface, the charge-neutral
scaling and the gradient expansion become identical, i.e., the gradient
expansion for the slowly-varying gas is simply a {\em special case} of
charge-neutral scaling.  To see that this is so, consider just the
kinetic energy as a density-functional.  Here the gradient expansion
is known out to 6th order\cite{M81}, and eventually the integrated quantity itself
diverges for atomic densities, due to the evanescent tail.  But
no such divergence occurs for extended systems with finite density everywhere\cite{M81,YPKZ97}.

\begin{figure}[htb]
\unitlength1cm
\begin{picture}(12,6.5)
\put(-6.2,-3){\makebox(12,6.5){
\includegraphics{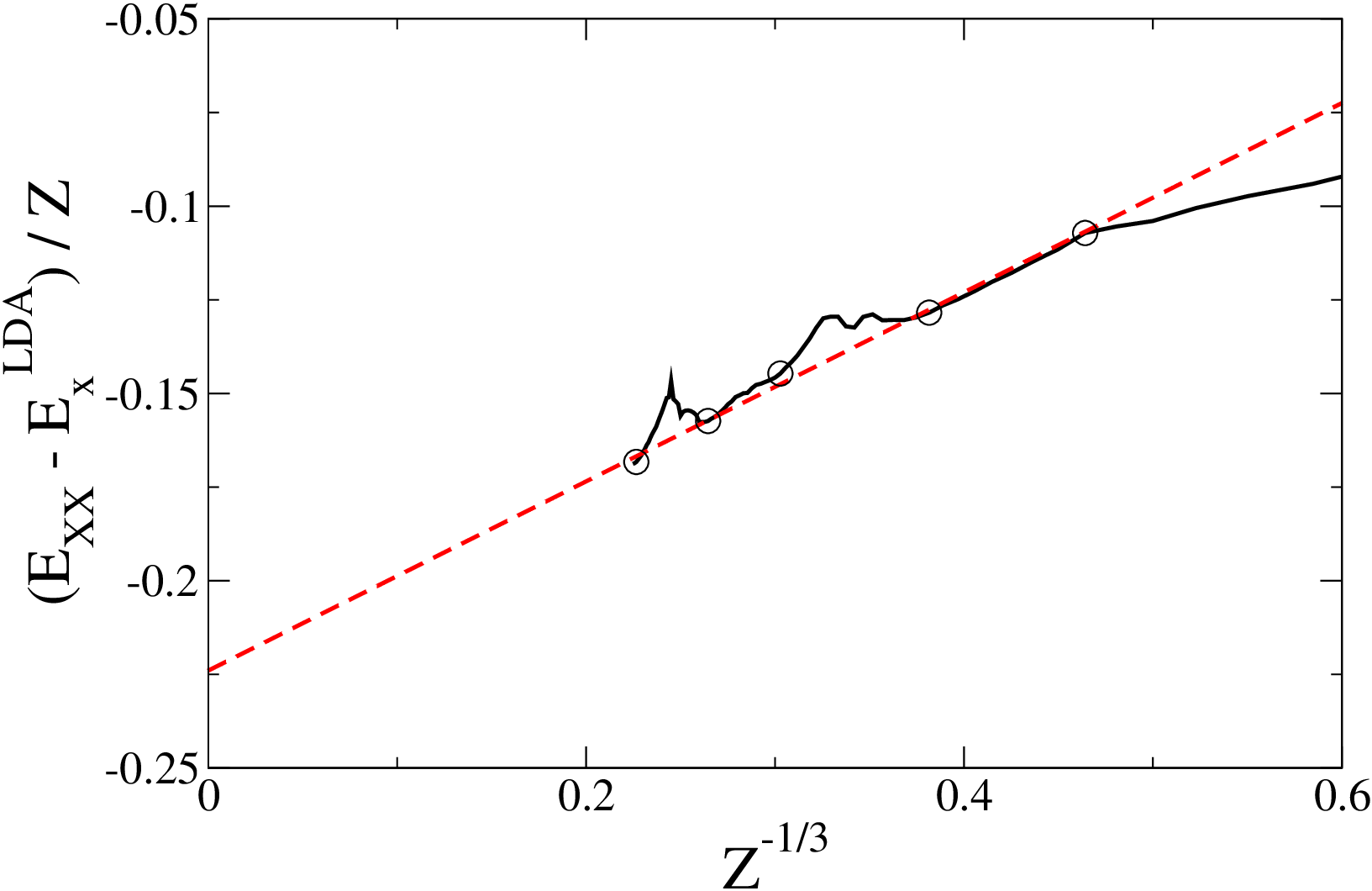}
}}
\end{picture}
\caption{We now find the next coefficient in the exchange asymptotic series. To minimize the error due to shell structure oscillations, the LDA exchange energy is subtracted from the exact exchange for each atom. As both give the  leading correction, their difference will then have $\Delta c Z$, $\Delta c = c_1 - c_1\LDA$, as the leading term in its asymptotic expansion, . The dashed line is the result of fitting to the noble gas atoms (circle symbols).}
\label{f:EXXmLDA}
\end{figure}

Thus CN scaling applies to all systems, but only becomes identical to 
the gradient expansion for slowly-varying bulk systems.  For the dominant
contribution, effectively the local Fermi wavelength becomes short on the
length scale on which the density is changing, so that the local approximation
applies, and yields the exact answer for this term.  Hence LDA reproduces
the AE0 terms, but GEA does {\em not} produce the leading corrections.
All this has been amply demonstrated for simple 1d model systems\cite{ELCB08},
and for the Kohn-Sham kinetic energy for atoms\cite{LCPB09}.

Here we apply the same reasoning for exchange.  The local approximation
becomes relatively exact as $\zeta\to \infty$, but the gradient expansion
does not reproduce the leading correction in the CN expansion.
Below, we use the simple reasoning of Ref. \onlinecite{PCSB06} to recover this leading
correction.  We perform a much more extensive calculation of the
asymptotic behavior, using methods developed in Ref. \onlinecite{LCPB09}.
We find, in agreement with Ref. \onlinecite{PCSB06}, that the leading correction for
atoms is about double that given by the gradient expansion, matching
quite closely that of B$88$ and of PBE.
Reversing this logic
for B$88$, we show that B$88$ may be more or less derived non-empirically
via the constraint that the approximation be AE1. If we enforce AE1 exactly, we find a slightly 
different value for $\beta$, and discuss the properties of the resulting functional, excogitated B$88$.

\begin{figure}[htb]
\unitlength1cm
\begin{picture}(12,6.5)
\put(-6.2,-3.3){\makebox(12,6.5){
\includegraphics{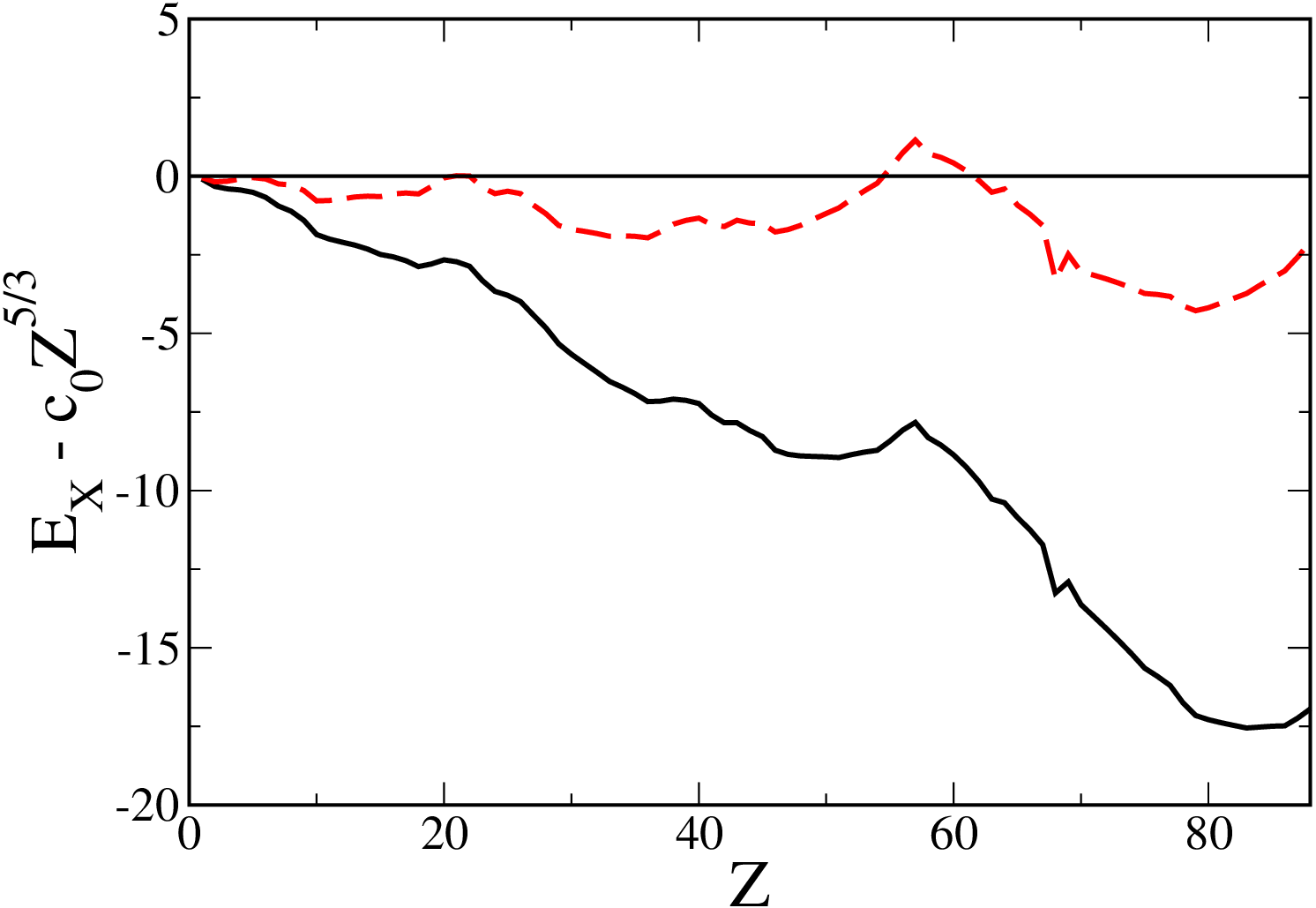}
}}
\end{picture}
\caption{In order to see that LDA does not significantly contribute to the higher orders of the exchange asymptotic series, we plot the difference between LDA and the leading term $c_0Z^{5/3}$ (dashed line), as a function of Z. The exact exchange value is also shown (solid line).}
\label{f:Exc0}
\end{figure}

\section{Extracting asymptotic coefficients}

Under the potential scaling of Eq. (\ref{vzeta}), any approximation for
the exchange energy that reduces to LDA in the uniform limit
has an expansion in $N$
like Eq. (\ref{ExSchwing}), with the same value for $c_0$. However the coefficient
$c_1$ depends on the particular approximation. 
Below we explain the procedure used to extract these coefficients.

As mentioned in the previous section, the OPMKS\cite{Engel} electronic structure code
is a fully numerical electronic structure code that has the ability to perform
optimized effective potential (OEP) calculations. We evaluate the various approximations
using atomic densities found with the OEP exact
exchange (XX) method. The densities found using this method will be extremely 
close to the exact densities despite the fact that correlation is missing. 
Moreover the effect of correlation will contribute at higher orders
in the asymptotic expansions of the energy than those we are interested in.
Thus EXX calculations are in principle sufficient for extracting
the coefficient we seek.

In Fig. \ref{f:EXX}, we plot $E\xx/Z^{5/3}$ vs $Z^{-1/3}$
where $E\xx$ is the exchange energy from the exact exchange calculation. 
Since the leading term in the asymptotic expansion of Eq. (\ref{ExSchwing}) is
$Z^{5/3}$, this procedure picks out the $c_0$
coefficient as a constant while all other terms are functions of $Z^{-1/3}$.
One can see that the curve in Fig. \ref{f:EXX} is heading towards the exact value of $c_0 = -0.2208$,
but it is difficult to extract higher coefficients due to oscillation of
the curve due to the shell structure.   

\begin{figure}[htb]
\unitlength1cm
\begin{picture}(12,6.5)
\put(-6.2,-3){\makebox(12,6.5){
\includegraphics{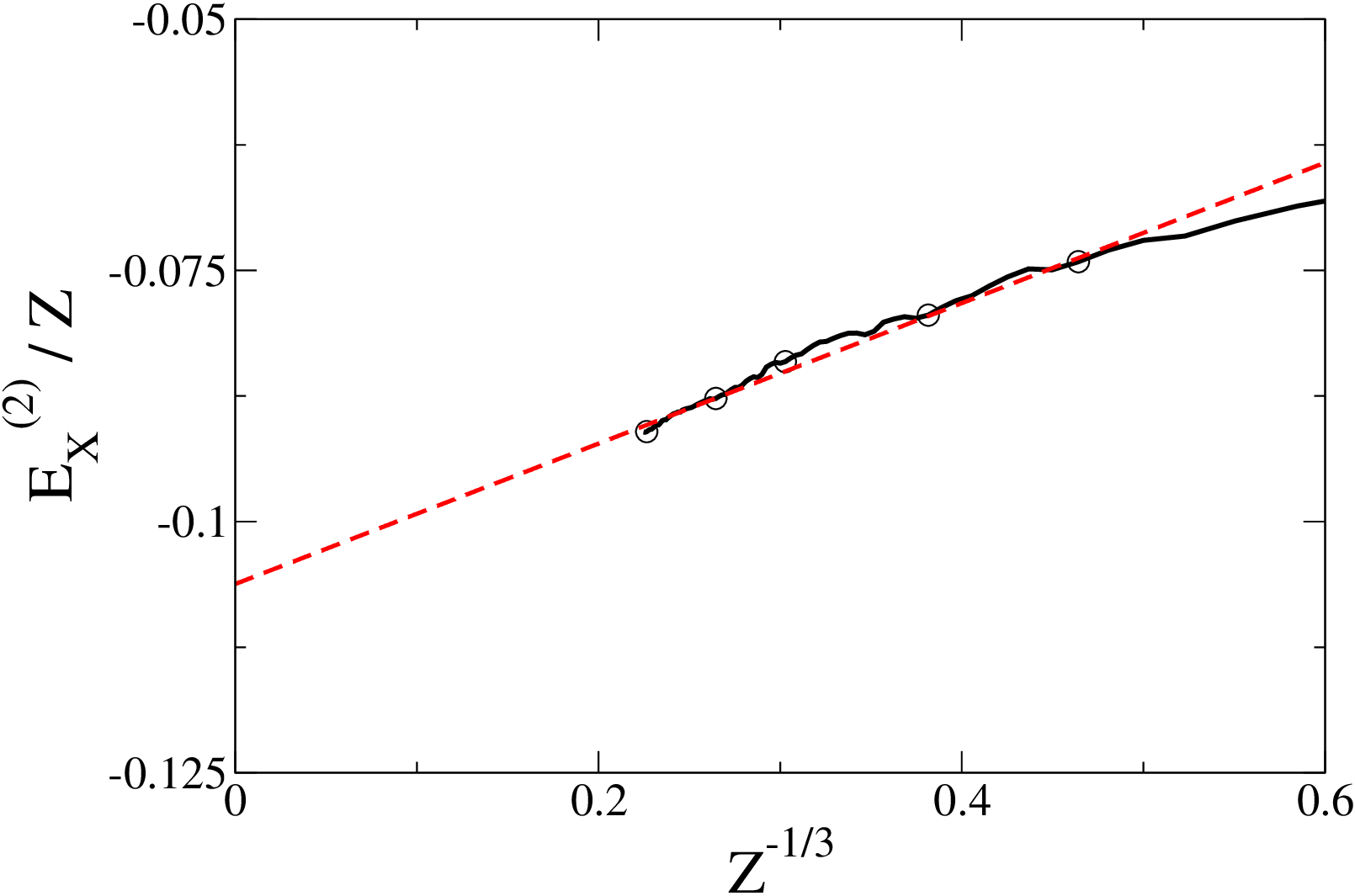}
}}
\end{picture}
\caption{We use the same procedure as in Fig. \ref{f:EXXmLDA} to find the $\Delta c$ coefficient for the gradient correction to LDA, $E\x^{(2)}[n]$, as defined in Eq. (\ref{Ex2}). The dashed line was fitted to the noble gas atoms (circle symbols).}
\label{f:dengea}
\end{figure}

To overcome this difficulty,
in Fig.\ref{f:LDA} we add the LDA curve to Fig.\ref{f:EXX}. It can be seen that
it too recovers the exact $c_0$ coefficient, but also clearly differs in higher orders
in the asymptotic expansion.  More usefully, we see that
the LDA curve mimics the oscillations shown by exact exchange,
so subtracting LDA from EXX will minimize this effect
and make the extraction of asymptotic coefficients more accurate.

In Fig. \ref{f:EXXmLDA}, we plot $(E\xx-E\x\LDA)/Z$ vs $Z^{-1/3}$ and find
that it behaves close to linearly.  There appears to be no $Z^{4/3}$ term in $E\x$.
Such a contribution was argued not to exist in Ref. \onlinecite{PCSB06}, but this was based
on CN scaling of the density and studying the behavior of the terms in the
gradient expansion.  That reasoning is insufficient, as the expansion should
be performed in terms of the potential, as described in Sec. \ref{s:pscal}.  But since the Scott correction (the $Z^2$
contribution to the total energy) comes from the core region, there is no reason
to expect an analogous contribution for exchange.  
In order to show this and also to precisely determine the $c_1$ coefficient,
one should use the techniques developed by Schwinger for deriving the Scott correction to the total energy\cite{S80},
but apply them to exchange.

To further reduce the remaining uncertainty due to shell structure oscillations, we choose simply
to use the noble gas atoms (excluding helium) for our fit.   The strongest deviations
from linearity come from the transition metals and lanthanides and actinides. We fit the difference
$(E\xx-E\x\LDA)/Z$ with a straight line in  $Z^{-1/3}$, and extrapolate to $Z\to\infty$,
finding 
$\Delta c = -0.2240$, where
$\Delta c = c_1 - c_1\LDA$, and 
\ben
\label{xxser}
E\xx \approx E\x\LDA - 0.2240\, Z +0.2467\, Z^{2/3} . 
\een
The coefficient of the last term is given by the slope of the dashed line in Fig. \ref{f:EXXmLDA}, although the meaning of this term is unclear in the presence of such strong
oscillating contributions.

\begin{figure}[htb]
\unitlength1cm
\begin{picture}(12,6.5)
\put(-6.2,-3){\makebox(12,6.5){
\includegraphics{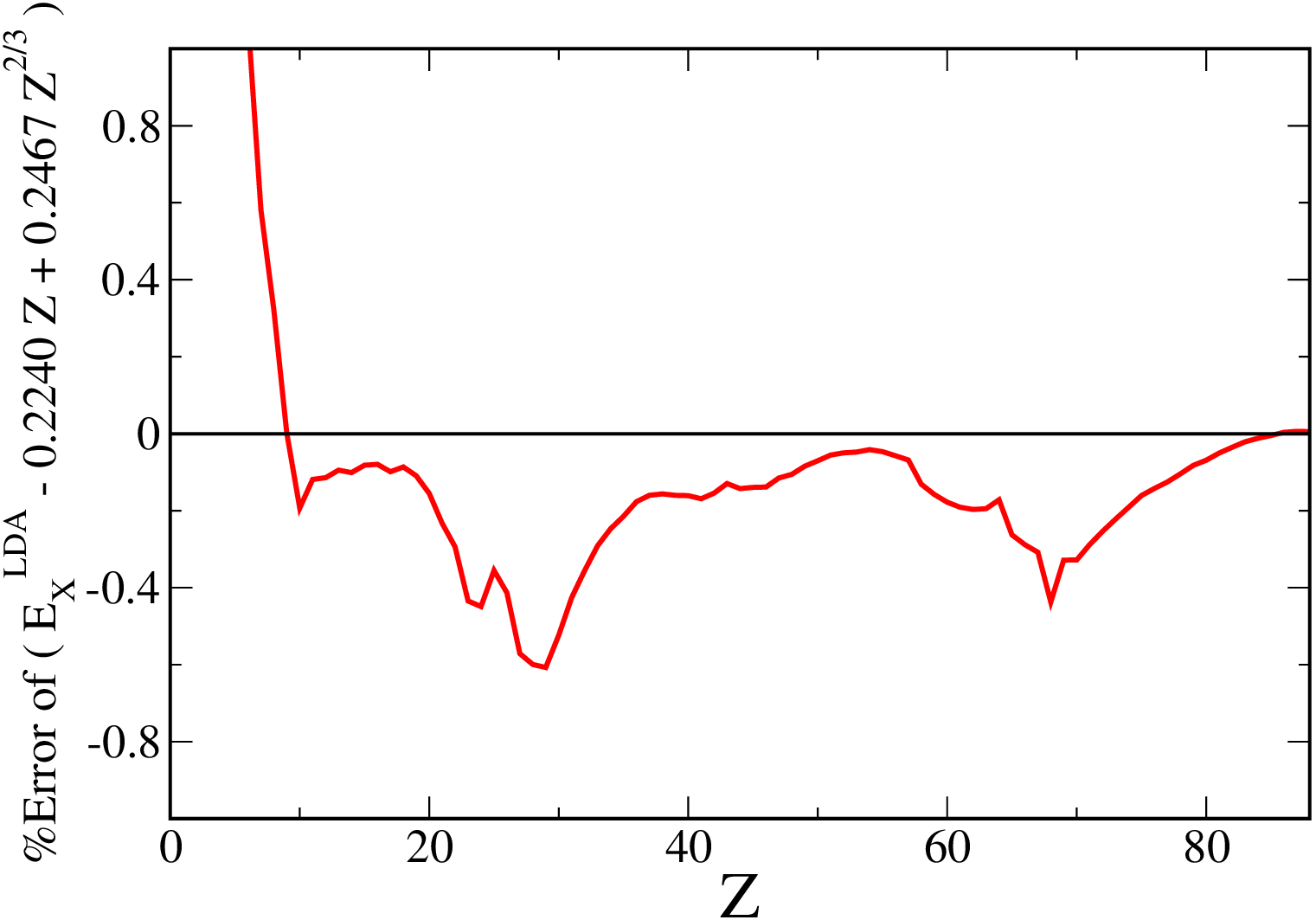}
}}
\end{picture}
\caption{The percentage error of the approximate asymptotic series given in Eq. (\ref{xxser}) is plotted as a function of $Z$. The error is remarkably low and demonstrates the power of these asymptotic series. }
\label{f:Exxser}
\end{figure}

If instead we used the alkaline earth atoms (excluding beryllium),
we find an almost identical value, $\Delta c = -0.2236$. 
If we use all elements with $Z>10$, we find a similar value,
$\Delta c = -0.2164$. 
If all elements from $Z=1$ to $88$ are used, 
we find $\Delta c = -0.1982$. 
In Ref. \onlinecite{PCSB06}, $\Delta c$ was found using noble gas atoms, 
except with the helium value included, and that method gave
a value of $\Delta c = -0.1978$. 
In our analysis, 
atoms with $Z<10$ are not used 
as they are not necessarily dominated by the asymptotic series.

Since LDA displays the shell oscillations that prevented us from fitting EXX
directly, the value of the LDA $c_1$ coefficient cannot be found exactly. 
But we estimate $0 \ge c_1\LDA \ge -0.04$, i.e., at least five times smaller in
magnitude than $\Delta c$.  In Fig \ref{f:Exc0}, we show $E\x - c_0\, Z^{5/3}$ as
a function of $Z$ for both the exact values and within LDA, demonstrating
that the linear contribution comes almost entirely from the beyond-LDA 
terms.

Finally we determine $\Delta c$ for GEA. In Fig. \ref{f:dengea}, 
we plot (E$\x\GEA-E\x\LDA)/Z$ vs Z$^{-1/3}$ in order to find $\Delta c = c_1\GEA - c_1\LDA$,
finding $\Delta c = -0.1062$.  This plot is much closer to linear than the previous one.
The leading corrections to LDA in the asymptotic expansion produce corrections to the
shell structure {\em beyond} those captured by LDA evaluated on the exact density
\cite{ELCB08,LCPB09}.   Although the smooth contribution can be partially
captured by GEA, there is almost no correction to the shell structure.
Just as for the kinetic energy\cite{LCPB09}, GEA yields a correction to the
smooth part that is about half of the accurate value.

\begin{figure}[htb]
\unitlength1cm
\begin{picture}(12,6.5)
\put(-6.2,-3){\makebox(12,6.5){
\includegraphics{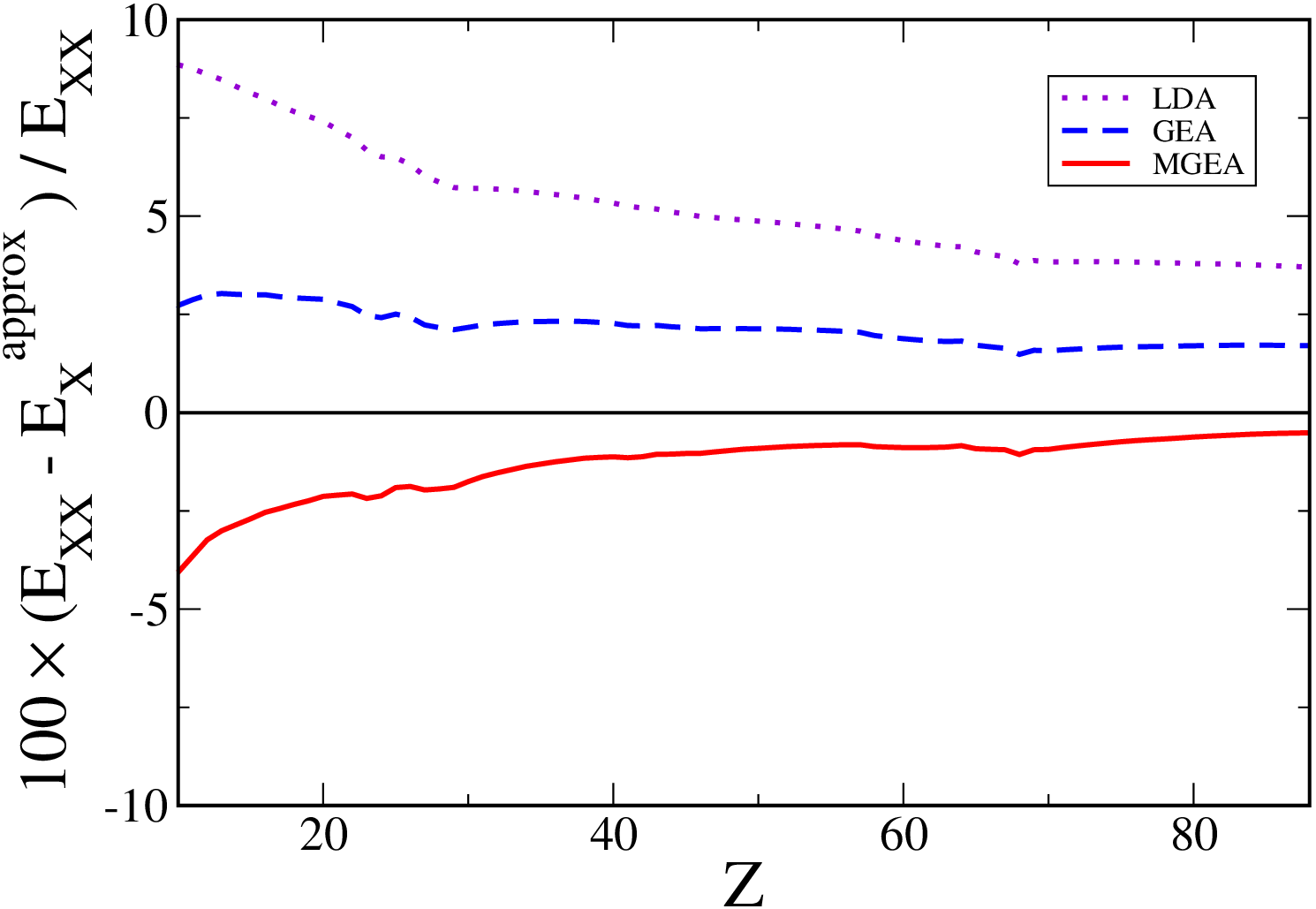}
}}
\end{picture}
\caption{The percentage error for LDA, GEA and the modified GEA (MGEA) exchange functionals for $Z>10$. The coefficient of $E\x^{(2)}$ is multiplied by $2.109$ in order to make the MGEA AE1 asymptotically exact.}
\label{f:errgea}
\end{figure}

To understand the power of these asymptotic expansions, we add the corrections
of Eq. (\ref{xxser}) to the LDA energies, and in Fig. \ref{f:Exxser} plot the percentage error relative to exact
exchange, as a function of $Z$.  For all but the second row of the periodic table,
the resulting error is below 0.5\% in magnitude, and typically of order 0.2\%.

\section{Generalized gradient approximations}

Generalized gradient expansions were designed to improve energetics over LDA for
electron systems of interest and relevance.  Early versions, such as PW91\cite{P91,BPW97}, were
tortured into reducing to the gradient expansion when the density is slowly varying.
But this was later given up, in both B88 and PBE exchange, which both
reduce to the gradient expansion form for slow variations, but with coefficients
much larger than that of the gradient expansion.

Our analysis explains why this must be so.  Regardless of its derivation, any modern
GGA for exchange is tested against the neutral atoms.  Any approximation that
cannot recover the right $c_1$ will be generally inaccurate for these energies,
and so discarded.  Thus any that become popular have already passed this test.  

\begin{table}
\caption{\label{tab:cs}$\Delta c = c_1 - c_1\LDA$ values for several different functionals.}
\begin{ruledtabular}
\begin{tabular}{cccccc}
 & E$\xx$ & LDA & GEA & B$88$ & PBE\\
\hline
$\Delta c$ & $-0.2240$ & -   & $-0.1062$ & $-0.2216$ & $-0.1946$
\end{tabular}
\end{ruledtabular}
\end{table} 

\begin{figure}[htb]
\unitlength1cm
\begin{picture}(12,6.5)
\put(-6.2,-3){\makebox(12,6.5){
\includegraphics{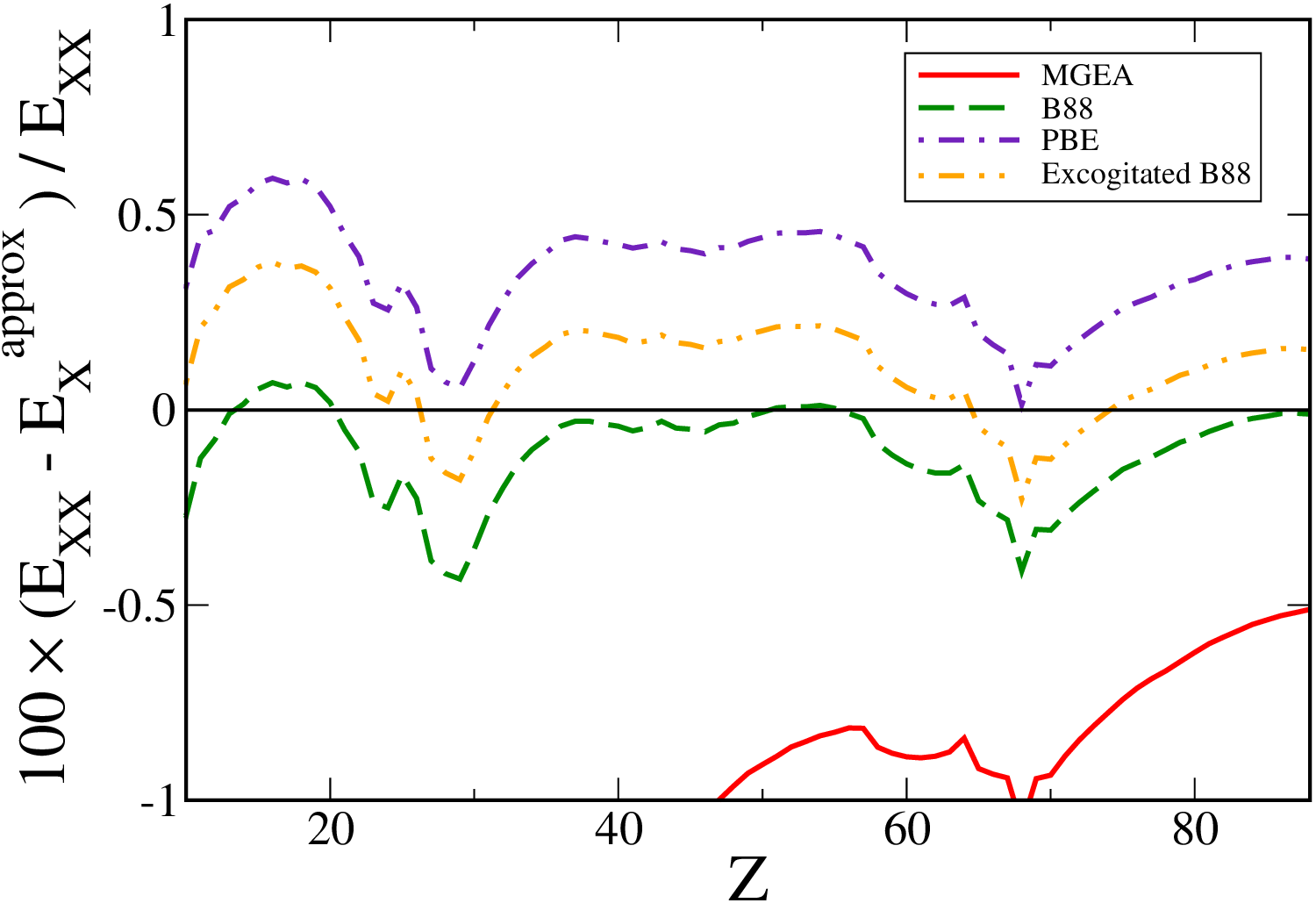}
}}
\end{picture}
\caption{We add to Fig. \ref{f:errgea} the results for B$88$, PBE, and the excogitated B$88$ functional, all evaluated on the OEP exact exchange densities.}
\label{f:errb88}
\end{figure}

In Table \ref{tab:cs}, we give the results for $\Delta c$ for several different functionals.
The same methodology was used in all extractions.
Both popular GGA's recover (at least approximately) the accurate value.
B88, designed specifically for molecular systems, is very close to the accurate
value.  PBE exchange is less so, but is also designed to bridge molecular and
solid-state systems. The PBE value is between that of GEA and B88, but much
closer to the latter than the former. Taking advantage of this insight, a
new variation on PBE, called PBEsol\cite{PRCV08}, restores the original gradient expansion,
thereby worsening atomization energies (total energy differences), but
improving many lattice constants of solids over PBE and LDA.

\subsection{Deriving the $\beta$ in B$88$}

The exchange energies found using GGAs such as B$88$ or PBE are generally
dominated by their gradient expansion
components for most chemically relevant densities. Thus, to make B$88$ AE1, it is
sufficient to impose this exact condition on just the $E\x^{(2)}[n]$ functional form. 
Since both B$88$ and the GEA are built on top of LDA, we can simple look at the $\Delta c$ values calculated above. If we set $\mu = 2.109\mu_{\sss AK} $ in Eq. (\ref{Ex2}), we multiply the $c_1$ coefficient of GEA by a factor of $2.109$, making it AE1. In Fig. \ref{f:errgea}, we name this functional MGEA for modified GEA and plot its percentage errors. The values for LDA and GEA are also shown. It can be seen that modifying GEA to be AE1 has greatly reduced the error.  

We now require that the B$88$ functional form, Eq. (\ref{B88}), reduce to this MGEA for small values of $x$. Using Eq. (\ref{mubeta}), this corresponds to using a value of $\beta = 0.0050$. Thus we have derived an excogitated B$88$ that is free of any empirical parameters. The actual value used in B$88$ is $\beta^{\sss B88} = 0.0053$ (for spin-polarized systems this becomes $0.0042$, which is the value given in Ref. (\onlinecite{B88})), so the values are very close. 
This is not surprising as fitting to Hartree-Fock exchange energies is an approximate way of demanding asymptotic exactness. Interestingly, the value quoted as the high-$Z$ asymptote in Ref. \onlinecite{B88} and found using Ref. \onlinecite{B86} is essentially the same as our value, but was evidently rejected in favor of a better fit.
In Fig. \ref{f:errb88}, we plot the percentage errors for B$88$, PBE, and the excogitated B$88$. As is typical for empirically-fitted functionals, B$88$ performs very well for systems close to the data set used in the fitted procedure. Although the error for PBE is higher than both B$88$ and the excogitated B$88$, it is systematic in its overestimation. As noted, PBE was designed to perform reasonably well for a wide range of systems, so again its behavior is not surprising. On this data set, the excogitated B$88$ was never going to do better than B$88$, although it remains to be seen how it performs for more complicated systems.

\section{conclusion}

We have carefully and systematically extracted the leading large-$Z$ correction
to the exchange energy of atoms.  Our results differ slightly from those of
Ref. \onlinecite{PCSB06} but yield the same qualitative conclusion, i.e., that the
gradient expansion yields an error of a factor of $2$ or more for this
coefficient.  We have clarified some of the reasoning, and applied it
more generally to any atom, molecule, or cluster.
By looking in detail at the exchange energy asymptotic series for neutral atoms, we have demonstrated the power of using such series for functional development. Requiring that the small gradient expansion of B$88$ capture the two leading coefficients of the asymptotic expansion is a method by which the unknown coefficient $\beta$ can be found. This gives a coefficient very close to the one actually used in B$88$ and thus is an {\it ex post facto} `derivation' of B$88$. Inserting our most accurate estimate for $\beta$ into the B$88$ form yields an excogitated B$88$.

We thank Eberhard Engel for the use of his atomic OPMKS code, and NSF CHE-0809859 .

\end{document}

%% file: dft.tex
\def\bea{\begin{eqnarray}}
\def\eea{\end{eqnarray}}
\def\ben{\begin{equation}}
\def\een{\end{equation}}
\def\benu{\begin{enumerate}}
\def\enu{\end{enumerate}}

\def\n{n}

\def\sss{\scriptscriptstyle\rm}

\def\1var{(\bx_1...\bx\N)}

\def\br{{\bf r}}

\def\bx{{\br t}}

\def\x{_{\sss X}}

\def\xc{_{\sss XC}}

\def\N{_{\sss N}}

\def\ext{_{\rm ext}}

\def\LDA{^{\rm LDA}}
\def\GEA{^{\rm GEA}}

\def\PBE{^{\rm PBE}}

\def\sph_int{ {\int d^3 r}}
